\documentclass[a4paper, 12pt]{article}

\usepackage[cp1251]{inputenc}
\usepackage{amssymb,amsmath}
\usepackage{graphicx}
\usepackage{dsfont}
\usepackage{color}
\usepackage{ulem}
\usepackage{verbatim}
\usepackage{epstopdf}
\usepackage{float}
\usepackage{cancel}
\usepackage{geometry}
\begin{document}

\begin{center}
{\bfseries\large Galaxy rotation curves in the $\mu$-deformation
 based approach to dark matter}

\vspace{1cm}

A.M.Gavrilik$^\sharp$, I.I.Kachurik$^\S$, M.V.Khelashvili$^\sharp$

\vspace{0.5cm}

$\phantom{x}^\sharp$ Bogolyubov Institute for Theoretical Physics,
NAS of Ukraine ({\it 14b, Metrolohichna Str., Kyiv 03143, Ukraine};\
\, e-mail:\,{\it omgavr@bitp.kiev.ua})

$\phantom{x}^\S$  Khmelnytski National University ({\it 11,
Instytuts’ka str., Khmelnytskyi, 29016, Ukraine})
   %11, Instytuts’ka str., Khmelnytskyi, 29016, Ukraine})

\end{center}

\setcounter{page}{1}%

\vspace{0.5cm}

\begin{center}
{\bfseries\large Abstract}
\end{center}

We elaborate further the $\mu$-deformation-based approach to
modeling dark matter, in addition to the earlier proposed use of
$\mu$-deformed thermodynamics.
 Herein, we construct $\mu$-deformed analogs of the Lane-Emden equation
 (for density profiles), and find their solutions.
  Using these, we plot the rotation curves for a number of galaxies.
 Different curves describing chosen galaxies are
 labeled by respective (differing) values of the deformation parameter $\mu$.
  As result, the use of $\mu$-deformation leads to
  improved agreement with observational data.
  For all the considered galaxies, the obtained rotation curves
(labeled by $\mu$) agree better with data, as compared to the well
known Bose-Einstein condensate model results of T. Harko. Besides,
for five of the eight cases of galaxies we find better picture for
rotation curves than the corresponding Navarro-Frenk-White (NFW)
curves. Possible physical meaning of the parameter $\mu$, basic for
this version of $\mu$-deformation, is briefly discussed.

\section{Introduction}

Model of dark matter as a Bose-Einstein condensate (BEC) of scalar
particles arose as an alternative to cold dark matter (CDM)
paradigm. It provides a possibility to resolve several tensions,
which CDM faces on the small scales, such as core-cusp problem and
the overabundance of the small-scale structures \cite{BEC}. We
should mention, however, that BEC model is not unique in this
aspect, and models like warm dark matter (warm DM), self-interacting
DM are also able to solve CDM problems on the small scales.

The BEC model considers the ultralight DM galaxy halo as a stable
``core'' solution of nonlinear Schr$\rm\ddot{o}$dinger (or
Gross-Pitaevsky) equation, with classical Poisson equation for
gravitational potential of DM halo, surrounded by DM envelope that
mimics CDM halo on the larger distances from the center of
galaxy~\cite{HOW}.
 Analysis of luminous matter kinematics in galaxies, like say
 dwarf spheroidal galaxies kinematics, indicates that coherent
 state core can represent all required DM in the dwarf galaxies,
 but only some smaller fraction of DM in bigger %\footnotetext[1]
           %{This work is based on the results presented at the
           %XI Bolyai-Gauss-Lobachevsky (BGL-2019) Conference: Non-Eucli\-de\-an,
           %Noncommutative Geometry and Quantum Physics.}
galaxies \cite{BEC_dSph}.

The DM particles within the considered model are ultra-light scalar
ones with mass  $m \!\sim\! (1\!-\!10)\cdot10^{-22} eV$ that is in a
good agreement with most observations (except only Lyman-$\alpha$
forest). Scalar particles with such extremely small
 mass\footnote{Note that in some works, see e.g. \cite{Kun2018},
 the role of Bose-condensed particles of dark matter is played by
 gravitons of tiny mass, bound from above by $m_g \sim 10^{-26} eV$.}
 %%%%%%%%%%%%%%%%%%%%%%%%%%%%%
can be considered as axion-like particle,  so that its mass is
protected against radiative correction by nonexact shift symmetry
$\phi \to \phi+C$. It is usually based on a free scalar field with
$\phi^4$ self-interaction potential~\cite{HOW}.

There exist however different extensions of this model, some of
which introduce more complex (than $\phi^4$) self-interacting
potential.
 Another interesting way is to consider non-minimal coupling
 of condensate to the gravity e.g. through potentials
 $G_{\mu\nu}\nabla^{\mu}\phi\nabla^{\nu}\phi $ and
 $\nabla_{\mu}\phi\nabla^{\mu}\phi R $ (here $G_{\mu\nu}$ -- Einstein
tensor, $R$ -- scalar curvature) \cite{Bet2014, Bet2011}.

A completely different direction constitute the DM models based on
non-standard statistics, like a condensate  of particles obeying
infinite statistics \cite{Inf}, and also our preceding work in which
we have proposed the model of dark matter viewed as a condensate of
a gas obeying $\mu$-deformed thermostatistics~\cite{GKKhN}.

\section{Bose Condensate Dark Matter Model: %}\subsection{
Gross-Pitaevsky and Lane-Emden equations}

BEC DM model suggests that DM consists of ultralight bosons of the
mass $10^{-22} eV$,  so its de Broglie wavelength is of astronomical
scale (kpc). Within this model, galaxy DM halo is represented by
halo of such particles, most of that are in the ground state, thus
forming a non-relativistic self-gravitating Bose-Einstein
condensate. If only particles in the ground state are taken into
account, such condensate halo can be described by the
Gross-Pitaevsky equation.
 Here we give brief overview of BEC halo description through
 Gross-Pitaevsky equation (see \cite{Harko} for more detailed discussion)
\begin{equation}   \label{Gross-Pit}
-\frac{\hbar^2}{2m}\nabla^2\Psi(\vec r) + V(\vec r)\Psi(\vec r) +
\frac{4\pi\hbar^2a}{m}|\Psi(\vec r)|^2 \Psi(\vec r) =
\tilde{\mu}\Psi(\vec r) .
\end{equation}
Herein $\Psi(\vec r)$ is wave function of the ground state, $
\tilde{\mu}$ denotes  chemical potential, a term $ \propto
|\Psi(\vec r)|^2 \Psi(\vec r)$ is responsible for condensate
particles self-scattering and $V(\vec r)$ -- represents any external
potential, that in the considered case will be Newtonian
gravitational potential of DM halo that obeys Poisson equation:
\begin{equation}                 \label{Poisson}
 \nabla^2 V(\vec r) = 4\pi G\rho(\vec r) .
\end{equation}

Particles in the condensate state are suggested to be
non-relativistic with almost zero temperature, so Thomas-Fermi
approximation, wherein kinetic term of the equation is neglected, is
applicable here. The corresponding equation can be rewritten by
introducing particles density
\[
\rho(\vec r) = m |\Psi(\vec r)|^2
\]
instead of wave-function, and takes a simpler form:
\[
\rho(\vec r) = \frac{m^2}{4\pi\hbar^2a}\left( {\tilde\mu} - mV(\vec
r)\right).
\]
Then, similarly to \cite{Harko}, we apply Laplace operator to the
latter equation and use \eqref{Poisson}, that yields
\begin{equation}                       \label{LE}
\Delta_r \rho(r) + k^2\rho(r) = 0  \hspace{8mm} {\rm with}
\hspace{6mm} k^2\equiv\frac{G m^3 }{\hbar^2 a} \,
\end{equation}
where $a$ is scattering length, i.e. the Lane-Emden equation with
polytropic index $n=1$. This equation admits simple analytical
solution for DM halo density within BEC DM model:
\begin{equation}                       \label{BEC-density}
 \rho(r) = \rho_c \frac{\sin kr}{kr}.
\end{equation}
The solution contains two free parameters $\rho_c = \rho(0)$ --
density in the DM halo center and the parameter $k$, which is
related to the total halo radius $R$ as $k = \pi/R$.

It should be mentioned that this solution takes into account only
particles in the ground state, however in a more realistic
description other states should also be considered. It is known that
ultra-light DM halo consists of static core (which allows to solve
core/cusp problem) surrounded by envelope that on the larger scales
mimic CDM behavior \cite{HOW}.
 The solution (\ref{BEC-density}) is responsible only for a core part
 of a galaxy DM halo, so we expect
that it will provide good explanation of observations on the scales
smaller than the size of core and at the same time will probably
meet some tension with observations of more distant regions of
galaxies.

It is also worth to mention that the same approximate equation can
be obtained from the Klein-Gordon equation for a scalar field, that
makes these models closely connected.

\section{Deformation of Lane-Emden equation}

Since a $\mu$-deformed analog of the Gross-Pitaevsky equation at
present is not available (and constitutes a nontrivial problem),
here we concentrate on performing $\mu$-deformation of the
Lane-Emden equation.

\subsection{Elements of $\mu$-calculus}

As our approach exploits so-called $\mu$-calculus, let us first
sketch it briefly (more detailed introduction to the $\mu$-calculus
and applying it to deformed models is given in~\cite{GKKhN,RKG}).
 Basic notion of this approach is the $\mu$-bracket (with $X$ being a
number or an operator):
$$
[X]_{\mu} = \frac{X}{1+\mu X}, \hspace{12mm}
 \mu\geq 0\, .
$$
Obviously, $[X]_{\mu}\to X$ if $\mu\to 0$.
 Using $\mu$-bracket we define the $\mu$-deformed
 (or $\mu$-) derivative such that
\begin{equation}                \label{mu-deriv}
\mathcal{D}_x^{(\mu)} x^n = [n]_{\mu} x^{n-1}\,.
\end{equation}
 The $\mu$-derivative does not satisfy\footnote{It is however %nevertheless
 possible to introduce a $\mu$-analog of Leibnitz rule
though its definition is rather nontrivial~\cite{RKG}.}
 Leibnitz rule, i.e.
$$
\mathcal{D}_{\mu} (f\cdot g) \ne f\cdot(\mathcal{D}_{\mu}g) +
(\mathcal{D}_{\mu}f) \cdot g .
$$
Note that the above action (\ref{mu-deriv}) implies the following
presentation of the $\mu$-derivative in terms of usual derivative in
the form of formal power series
\begin{equation}                            \label{mu-derv}
%\[
\mathcal{D}_x^{(\mu)} \equiv \left[ \frac{d}{dx} \right]_\mu=
\frac{\frac{d}{dx}}{1+\mu\frac{d}{dx}}
%\]
%\vspace{-2mm}
  \hspace{1mm} = \frac{d}{dx}\Bigl(1-\mu\frac{d}{dx}+\mu^2\frac{d}{dx}\frac{d}{dx}-...\Bigr)
 \end{equation}
that incorporates all higher orders of the derivative
$\frac{d}{dx}$.  This fact is of basic importance.

Now, one can introduce deformation in the theory of interest merely
by substituting each derivative in equations by its deformed analog.

The $\mu$-bracket is used in deformed versions of known functions,
say, the $\mu$-deformed exponent
$$
\exp_{\mu} x = \sum\limits_{n=0}^{\infty} \frac{x^n}{[n]_{\mu}!}
$$
allows to define deformed sine and cosine functions:
$$
\sin_{\mu} x = \frac{1}{2i}\left(\exp_{\mu} (ix) - \exp_{\mu}
(-ix)\right)
%$$
% \vspace{-7mm}
% $$  \hspace{-18mm}
 = \sum\limits_{n=0}^{\infty} (-1)^n\frac{x^{2n+1}}{[2n+1]_{\mu}!} ,
$$
$$
\cos_{\mu} (x) = \frac{1}{2}\left(\exp_{\mu} (ix) + \exp_{\mu}
(-ix)\right)
%$$
% \vspace{-7mm}
% $$  \hspace{-18mm}
= \sum\limits_{n=0}^{\infty} (-1)^n \frac{x^{2n}}{[2n]_{\mu}!}
$$
where the $\mu$-factorial means the product
 $[m]_{\mu}! = [m]_{\mu}\,[m\!-\!1]_{\mu} \ldots [2]_{\mu}\,[1]_{\mu}.
 $ \
 The deformed harmonic functions will arise in the next sections of the
paper. We have to stress the fact that the familiar differential
relations between harmonic functions (involving usual derivative)
are not valid in the $\mu$-deformed case:
  $\frac{d}{dx}\sin_{\mu}(x) \ne \cos_{\mu} (x)$.
 However, the deformed counterpart which uses
 the $\mu$-derivative does hold, namely
$$
 \mathcal{D}^{\mu}_{x} \sin_{\mu} (x) = \cos_{\mu} (x) \, .
$$
It is clear that the deformed analog of derivative and deformed
functions should reduce to their non-deformed versions in the
limiting case of $\mu \to 0$. It is a simple matter to restore
non-deformed versions of equations of underlying theory at any step
of analysis.

\subsection{Deforming Laplacian in the LE equation}

As already mentioned, DM halo density in the BEC DM model could be
approximately described by the Lane-Emden (LE) equation with
polytropic index $ n = 1 $:
\begin{equation}
\Delta_r \rho(r) + k^2\rho(r) = 0,
\label{SphLE}
\end{equation}
wherein $\Delta_r$ is the radial (thus 1-dimensional) part of the
spherical Laplace operator, namely
$$
\Delta_r f(r) = \frac{1}{r^2}\frac{d}{dr}\left(r^2 \frac{d}{dr} f(r)\right).
$$
The latter is also equal to
$$
\Delta_r f(r) = f''(r) + \frac{2}{r}f(r)\, ,
$$
and the LE equation can be written in its more familiar form
\begin{equation}
\left(\frac{d^2}{dr^2} + \frac{2}{r}\frac{d}{dr} +k^2 \right) \rho(r) = 0.
\label{LE_usual}
\end{equation}

In order to deform, we first take LE equation in the initial form
\eqref{SphLE} as a  starting point. Let's introduce the deformation
in the equation by replacing derivative with respect to $r$ by its
$\mu$-deformed analog, to get a $\mu$-deformed analog of LE
equation:
$$
\frac{1}{r^2}\mathcal{D}^{\mu}_r\left(r^2 \mathcal{D}^{\mu}_r
\rho(r)\right) + k^2 \rho(r) = 0.
$$
The equation can be %easily
rewritten through a dimensionless variable  $ x = kr $, and as
result we obtain
\begin{equation}
\frac{1}{x^2}\mathcal{D}^{\mu}_x\left(x^2 \mathcal{D}^{\mu}_x
\rho(x)\right)  + \rho(x) = 0. \label{deformLE}
\end{equation}
 %In addition
We adopt the same initial condition which was valid for
the solution of original equation:
$$
\rho(0) = \rho_c,   \hspace{10mm}  %$$  $$
\rho'(0) = 0 .
$$
In the last relation usual or $\mu$-deformed differentiation could
be applied.  Luckily, as it will become clear later, this does not
affect the results.

We are looking for a solution of eq.(8) in the form
$$
\rho(x) = \rho_c \sum\limits_{n=0}^{\infty} a_n x^n.
$$
 The operators in the equation act on the series as
$$
\mathcal{D}^{\mu}_x \cdot \, \rho(x)
 =  %\quad \to \quad \rho_c
\sum\limits_{n=1}^{\infty} a_n [n]_{\mu} x^{n-1} \, ,
$$
\vspace{-3mm}
$$
 \mathcal{D}^{\mu}_x \cdot \, \left(x^2 \, \rho(x)\right)
  =   %\quad \to \quad \rho_c
\sum\limits_{n=1}^{\infty} a_n [n]_{\mu}[n+1]_{\mu} x^{n}  \, .
$$
\vspace{-2mm} Then from the $\mu$-LE equation we infer
\begin{equation}                                  \label{LE-last}
\sum\limits_{n=1}^{\infty} a_n [n]_{\mu}[n+1]_{\mu} x^{n-2} +
\sum\limits_{n=0}^{\infty} a_n x^n = 0 \, ,
\end{equation}
and the initial conditions imply
$$
\rho(0) = \rho_c  = \rho_c a_0 \quad \to \quad a_0 = 1,
$$
\vspace{-5mm}
$$
\rho'(0) = 0 = \rho_c a_1 \quad \to \quad a_1 = 0.
$$
By changing summation limits we have
$$
\sum\limits_{n=0}^{\infty} a_{n+2} [n+2]_{\mu}[n+3]_{\mu} x^n +
\sum\limits_{n=0}^{\infty} a_n x^n = 0\, ,
$$
and due to this, can relate the coefficients:
$$
a_{n+2} = - a_n\frac{1}{[n+2]_{\mu}[n+3]_{\mu}}, \ \ \  n = 2m .
$$
That implies
 \vspace{-3mm}
$$
a_{2n} = (-1)^n \frac{1}{[2n]_{\mu}[2n+1]_{\mu}} \cdot ... \cdot \frac{1}{[3]_{\mu}[2]_{\mu}}
%$$
%or
  %\vspace{-5mm}
%$$  %a_{2n}   %%   \hspace{-15mm}
   = (-1)^n \frac{1}{[2n+1]_{\mu}!} [1]_{\mu} \, .
$$
The solution then takes the form
$$
\rho(kr)\! =\! \rho_c \, [1]_{\mu} \! \sum\limits_{n=0}^{\infty}
(-1)^n \frac{(kr)^{2n}}{[2n+1]_{\mu}!} = \rho_c \, [1]_{\mu} \,
\frac{\sin_{\mu} (kr)}{kr} .
$$
Thus, denoting $\rho_0 = \rho_c \,[1]_{\mu} $,
 we obtain
 \begin{equation}
\rho(kr) = \rho_0 \, \frac{\sin_{\mu} (kr)}{kr}
\label{Ldeform-density}
\end{equation}
as our main result for the DM density distribution.

\subsection{Deforming  derivatives in the LE equation}

As it is known for deformed models, different types of equation
deformation can be proposed.  Here we will present a different
version of $\mu$-deformed LE equation. We start with the LE equation
of polytropic index $n=1$ in its most common form \eqref{LE_usual},
and introduce deformation in the equation by replacing all spatial
derivatives  $d/dr$ with its deformed analog $\mathcal{D}^{\mu}_r$:
\begin{equation}                                \label{muLE-2}
\left(\mathcal{D}^{\mu}_r \mathcal{D}^{\mu}_r +
\frac{2}{r}\mathcal{D}^{\mu}_r +k^2 \right) \rho(r) = 0 \, .
\end{equation} Again we are looking for solution being power series:
$$
\rho(x) = \rho_c \sum\limits_{n=0}^{\infty} a_n x^n.
$$
Substituting this in the equation we have
$$
\ \ \ \sum\limits_{n=0}^{\infty} a_{n+2} [n+2]_{\mu}[n+1]_{\mu} x^n
+ 2 \sum\limits_{n=0}^{\infty} a_{n+1} [n+1]_{\mu} x^{n-1}
%$$
%  \vspace{-5mm}
 % $$
   %    \hspace{-50mm}
        + \sum\limits_{n=0}^{\infty} a_n x^n = 0
$$
Using same initial conditions
$\rho(0)  %%%& hspace{8mm}
= \rho_c,\   \rho'(0) %%%&
= 0 , $ after similar steps as above we obtain the
result\footnote{There exists another form of deformed LE equation
also possessing solution (11) -- with 1st term same as in
eq.(\ref{muLE-2}), but 2nd and 3rd terms multiplied by certain %respective
functions (of $k, r, \mu$). That will be explored elsewhere.}:
\begin{equation}                          \label{Ddeform-density}
\rho(r) = \rho_c\sum\limits_{n=0}^{\infty}
 (-1)^n\frac{(kr)^{2n}}{\prod\limits_{l=1}^{n}[2l]_{\mu}([2l-1]_{\mu}+2)}\, .
\end{equation}

\section{Galaxy Rotation Curves}

Now let us confront the predictions of our model with available
observational data.
 We analyze rotation curves of low surface brightness (LSB) galaxies,
 as kinematics of luminous matter in the galaxy depends on the density
 distribution within the galaxy. We have chosen those eight LSB galaxies
 which were analyzed by T.~Harko, in order to compare the models.
 Since these are DM-dominated, we neglect the gravitational contribution
 of baryonic matter.

 In the case of disk galaxies where trajectories of stars and clouds
 of gas could be assumed circular with a good accuracy, one can apply simple relation
$$
mv(r)^2 = \frac{GM(r)m}{r}
$$
that arises from virial theorem's relation between kinetic and
potential  energy $2T = V$ for a stable system in the gravitational
potential $V \propto r^{-1}$. Then, velocity $v(r)$ on the circular
orbit $r$ could be expressed as
$$
v(r) = \sqrt{\frac{GM(r)}{r}}
$$
This equation provides as a tool for studying density distribution
of matter  in the galaxy through observed rotation curves. We will
neglect gravitational effect of luminous matter, thus in previous
equation taking into account $M(r)$ of only dark matter component.

To define velocity $v(r)$ on any orbit $r$ we calculate  total mass
within this orbit
$$
M(r) = 4\pi \int\limits_{0}^{r} \rho(r')r'^{2}dr'
$$
In accordance with deformed differential relations with radial
coordinate $r$, we have to use deformed (or $\mu$-)
integration\footnote{This means applying the operator
$(D^{(\mu)}_r)^{-1}$, inverse to the one defined in eqs.(5), (6).}.
 %%%
As final result we obtain the following expression for velocity on
 circular orbit using the ``Laplace-deformed'' density solution
\eqref{Ldeform-density}:
\begin{equation}                                    \label{Ldeform-rotcurve}
v(r) = \sqrt{\frac{4\pi G\rho_0}{k^2}\sum\limits_{n=0}^{\infty}
\frac{(-1)^n(kr)^{2n+2}}{[2n+1]_{\mu}!\,[2n+3]_{\mu}}}\, .
\end{equation}
With the notation
$$
A = \sqrt{\frac{4\pi G\rho_0}{k^2}},
$$
%Let us also present
the respective expression for rotation velocity based on the
``derivatives-deformation'' solution \eqref{Ddeform-density} reads:
\begin{equation}                                        \label{Ddeform-rotcurve}
v(r) = A\sqrt{\sum\limits_{n=0}^{\infty} \frac{(-1)^n
 (kr)^{2n+2}}{[2n+3]_{\mu}\prod\limits_{l=1}^{n}[2l]_{\mu}([2l-1]_{\mu}+2)}}\, .
\end{equation}

We perform list-square analysis for the same eight LSB galaxies that
were studied  in \cite{Harko} regarding the standard BEC DM model.
The observational rotation curves of these galaxies were taken from
\cite{THING, Oman, Swaters}.

 Below, in Table 1 and Figure 1 (solid lines plot our curves, dashed
grey/black are NFW/BEC curves), we present results of fitting eight
LSB galaxies rotation curves by theoretical rotation curves: within
BEC DM as follows from its DM density solution \eqref{BEC-density},
the $\mu$-deformed ``Laplace deformation'' rotation curve
\eqref{Ldeform-rotcurve}, and the Navarro-Frenk-White profile
\cite{Oman} for CDM i.e.
$$
\rho_{NFW}(r) =\frac{\rho_0}{\frac{r}{R_s}\left(1+\frac{r}{R_s}
\right)^2} \, .
$$
  The least $\chi^2$ value among three studied models for
each of  galaxies is denoted by the bold font.

\vspace{0.5cm}

\noindent\begin{tabular}[h!]{|l|rrrr||rrr||r|c|}
\hline
Galaxy &    \multicolumn{4}{c||}{   $\mu-$BEC }             & \multicolumn{3}{c||}{BEC}           & NFW       & $N_{points}$ \\
\hline
    &       $A(km/s)$ & $k(kpc^{-1})$ & $\mu$  & $\chi^2$ & $A(km/s)$   & $k(kpc^{-1})$ &   $\chi^2$ &  $\chi^2$ &    \\
\hline \hline
DDO 53 &        32.27     & 0.97      & 0.180  & {\bfseries 1.9}     & 33.16      & 1.09       & 5.4     &  7.8     & 18 \\
HO I &      35.51     & 1.27      & 0.151    & {\bfseries 40.2}    & 34.78      & 1.27        & 160.1        & 109.8     & 22 \\
HO II &     40.11     & 0.40      & 0.179    & 31.8    & 37.93      & 0.45        & 66.8     & {\bfseries 6.4}      & 35 \\
IC 2574 &   81.37     & 0.17      & 0.179    & {\bfseries 13.7}    & 73.9       & 0.24        &  14.2    & 117.2         & 50 \\
NGC 2366    &   64.68     & 0.37      & 0.178    & {\bfseries 41.2}    & 72.72      & 0.37        & 126.1    & 71.9  & 41 \\
M81dwB &    38.35     & 2.64      & 0.180    & {\bfseries 2.5}     & 38.07      & 3.15       & 7.9   & 4.1       & 13 \\
DDO 154 &   51.30     & 0.38      & 0.156    & 199.3   & 53.81      & 0.38        & 549.7    & {\bfseries 76.8}  & 61 \\
DDO 39 &        87.52     & 0.27      & 0.174    & 168.2   & 87.96      & 0.30        & 343.0        & {\bfseries 14.3}  & 21 \\
\hline
\end{tabular}

\vspace{0.1cm}

\section{Discussion and Concluding Remarks}
 We explored, in addition to $\mu$-thermodynamics used~\cite{GKKhN} for modeling
 dark matter, the related approach based on $\mu$-deformed spatial derivative.
  Two differing $\mu$-deformed analogs of the Lane-Emden equation were studied,
  and their solutions, describing density profiles of DM halo, found.
   This allowed us to obtain the plots for the rotation curves
   of a number of galaxies.
 The corresponding curves  for chosen galaxies involve
   differing values of the deformation parameter $\mu$.
 As seen, nice agreement due to the use of $\mu$-deformation is achieved:
   for all the considered galaxies, our results show noticeable improvement as compared
   to the BEC model results of~\cite{Harko}.
%%%%%%%%%
   Moreover, the used approach provides somewhat better picture (agreement) even
 with respect to the famous NFW~\cite{Oman} rotation curves, say in
 the five (of the eight) cases, i.e. the curves for the galaxies DDO 53, HO I,
 IC 2574, NGC 2366 and M81dwB.

The importance and strength of $\mu$-deformation stems from certain
non-locality due to usage of deformed spatial $\mu$-derivative
\eqref{mu-derv}, that is, of extended operator built with usual
derivative in its denominator (that means, all orders of the
derivative $\frac{d}{dr}$ are present).
   This feature resembles such well-known approach as
   nonlocal modifications of gravity, see e.g.~\cite{Deser,Hehl,Arraut,Mitra,Park}
   and references therein. There is rather popular
   viewpoint that nonlocal gravity theories are of importance
   for solving basic problems of cosmology -- that of dark energy and dark matter.

In view of the success of the $\mu$-deformation-based description,
let us briefly discuss possible physical sense of the
$\mu$-deformation modifying the (radial) spatial derivative, and the
very parameter $\mu$.
 Being very massive but relatively compact (from cosmological scales
 viewpoint) objects, the DM halos can modify (geometry of) ambient space,
 and the employed $\mu$-derivative takes effectively into account such
 modification, with $\mu$ measuring the extent of modification.
 This agrees with the noticed important feature:  if we calculate total mass
 of galaxy DM halo (with fixed proper radius), we find that the bigger is
 the mass of halo, the greater respective value of $\mu$ should be taken.

At last, let us note that similar conclusions can be drawn by
applying the formula (\ref{Ddeform-rotcurve}).

\vspace{0.2cm}

\noindent {\bf Acknowledgement}
\vspace{0.1cm}

\noindent This work was partially supported by the
Special Program, Project No. 0117U000240, of Department of Physics
and Astronomy of National Academy of Sciences of Ukraine.

\vspace{-0.1cm}

\renewcommand{\refname}

\newpage

\begin{figure*}[h!]
               \vspace{20mm}
{\hspace{24mm} DDO 53 \hspace{80mm} M81dwB} \\ %[2mm]
              \vspace{4mm}
 \hspace{6mm}\includegraphics[width=72mm]{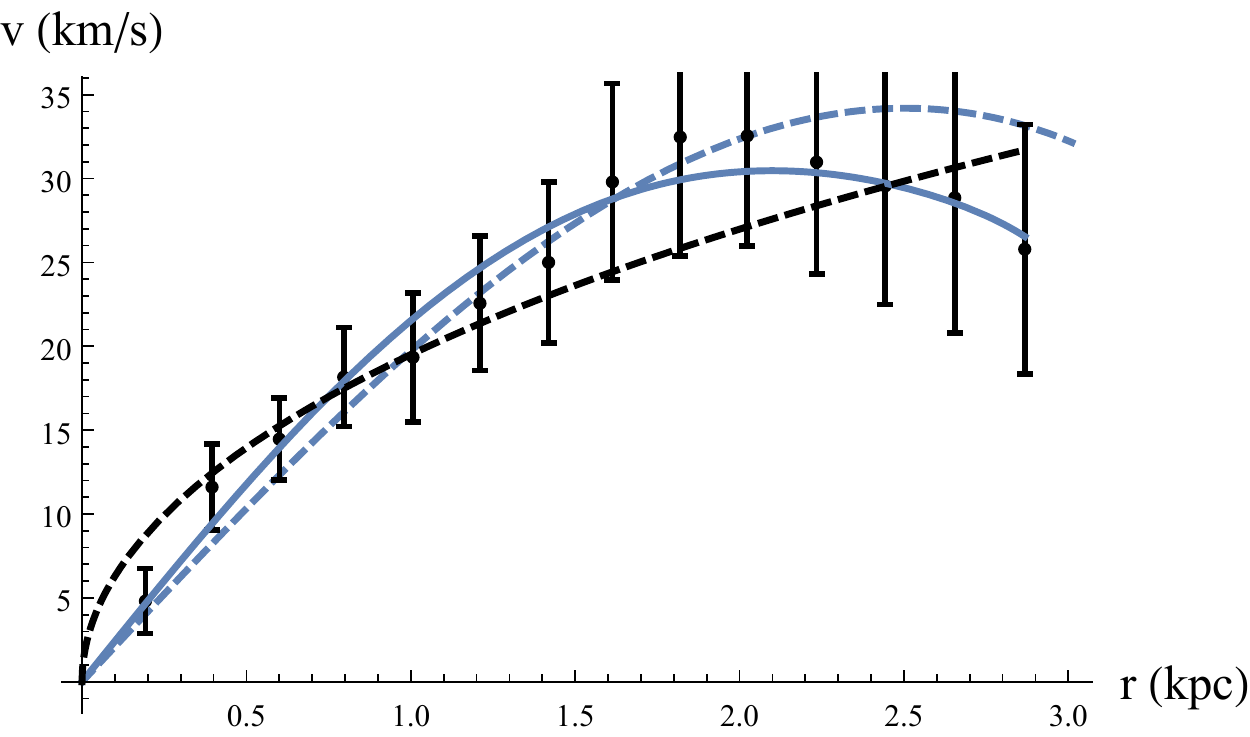} \hspace{25mm} \includegraphics[width=72mm]{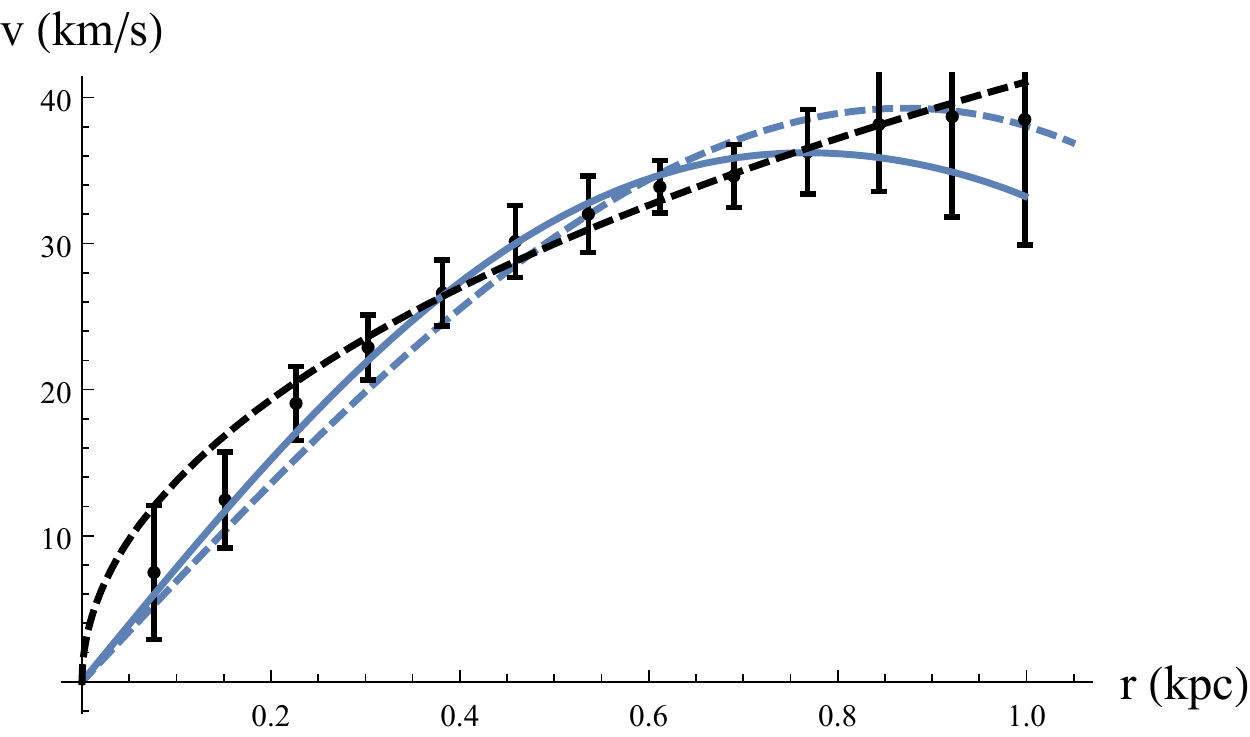} \\ %[2mm]
\vspace{0.5cm} \\  %[2mm]
   {\color{white} .} \hspace{22mm} HO I \hspace{85mm} HO II   \\ %[2mm]
              \vspace{1mm}
 \hspace{6mm} \includegraphics[width=72mm]{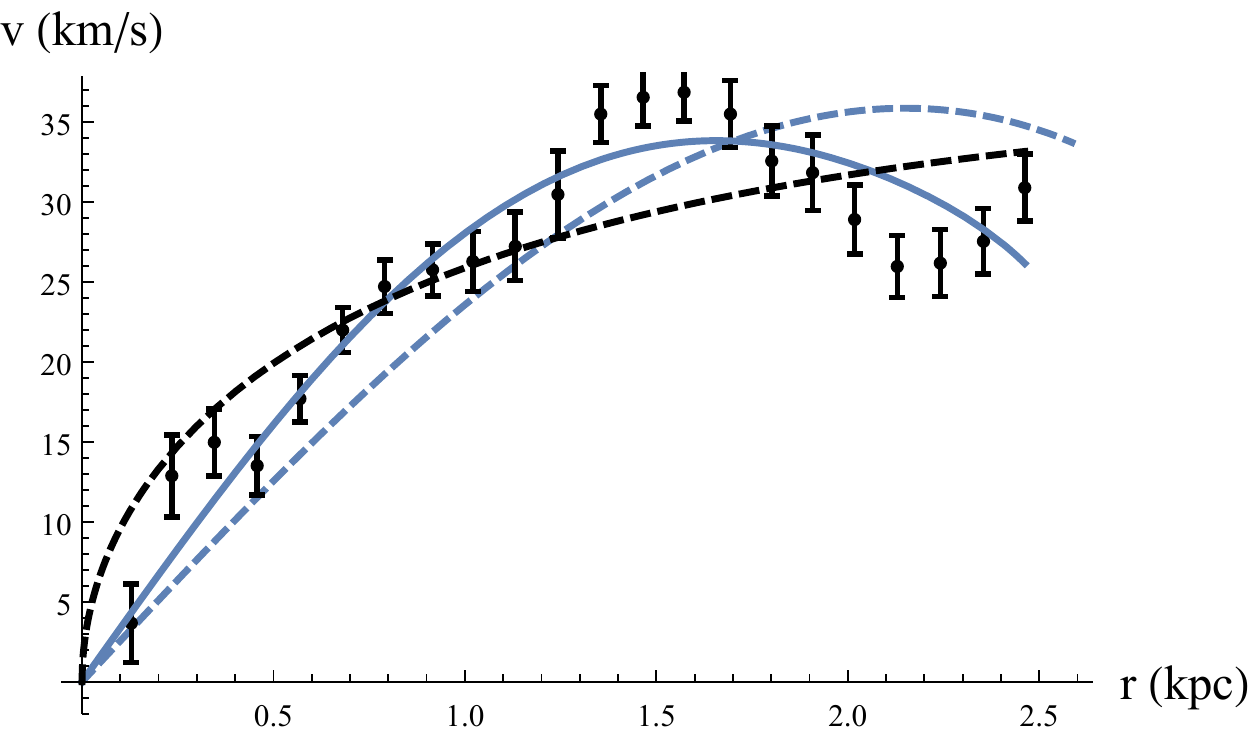}  \hspace{25mm}  \includegraphics[width=72mm]{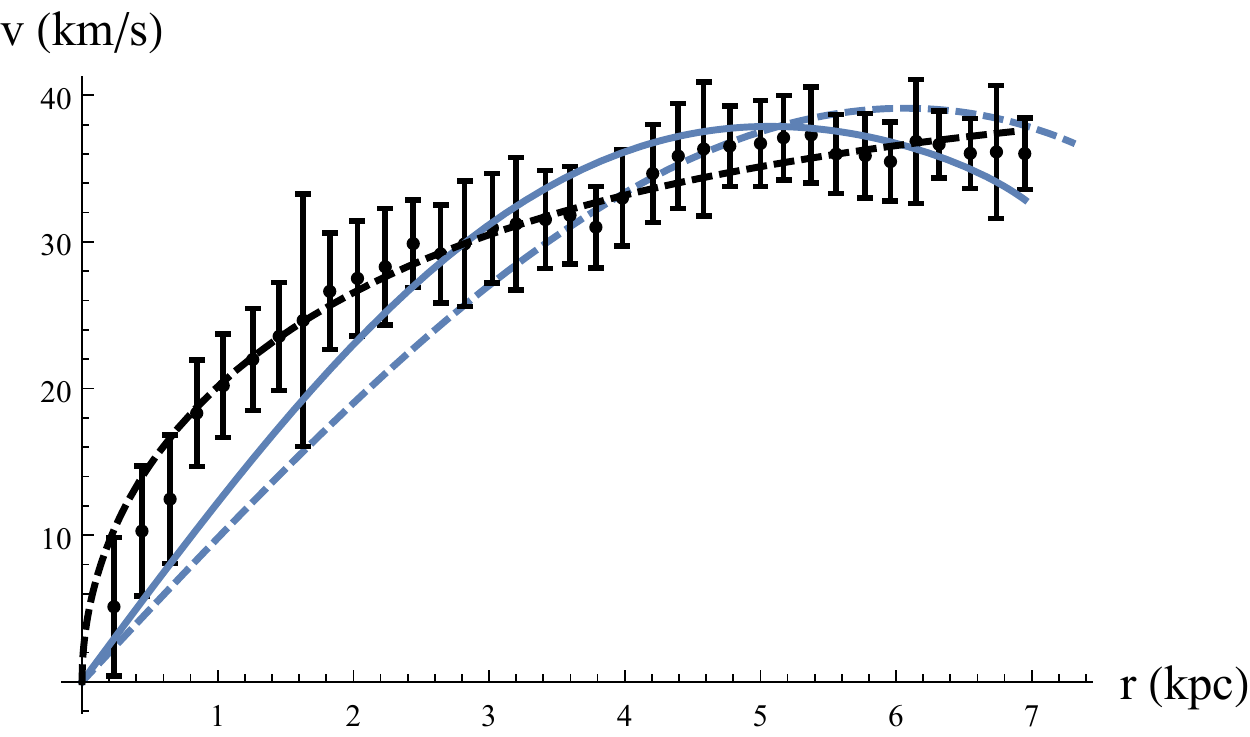} \\ %[2mm]
\vspace{0.5cm} \\  %[2mm]
    {\color{white} .} {\hspace{24mm} IC 2574 \hspace{75mm} NGC 2366 } \\ %[2mm]
                \vspace{2mm}
  \hspace{6mm} \includegraphics[width=72mm]{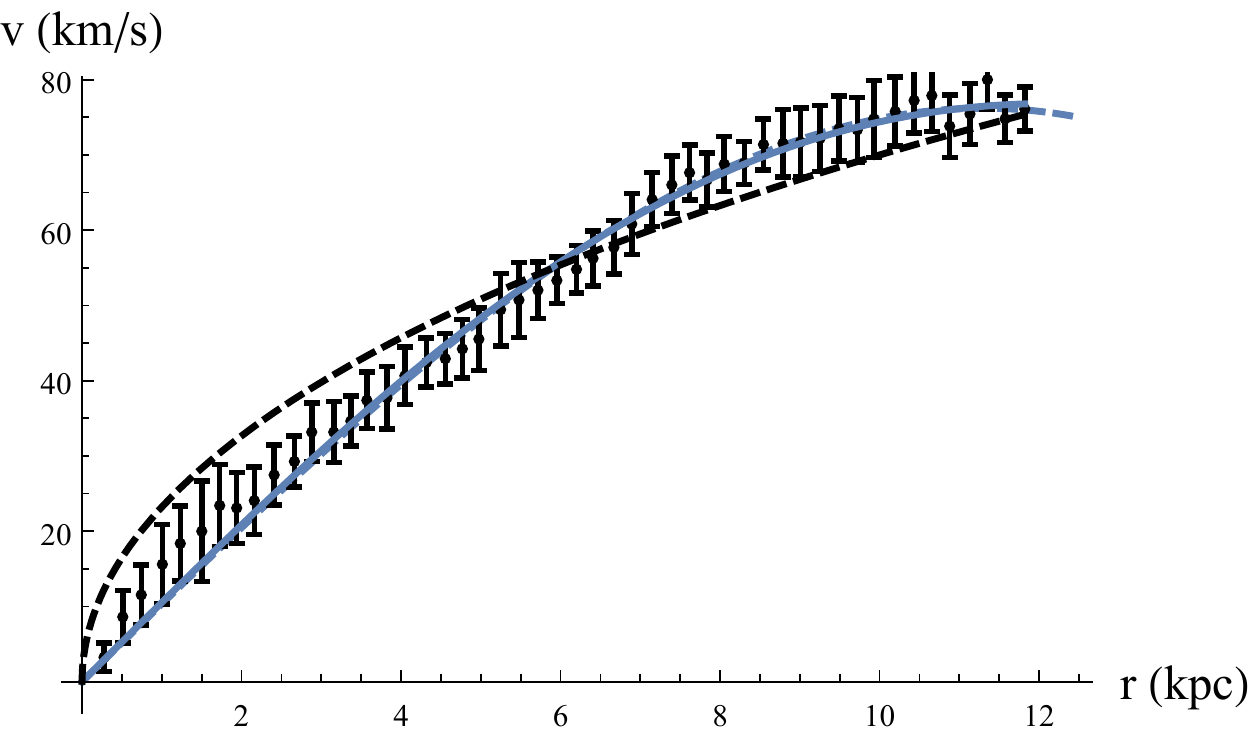} \hspace{25mm} \includegraphics[width=72mm]{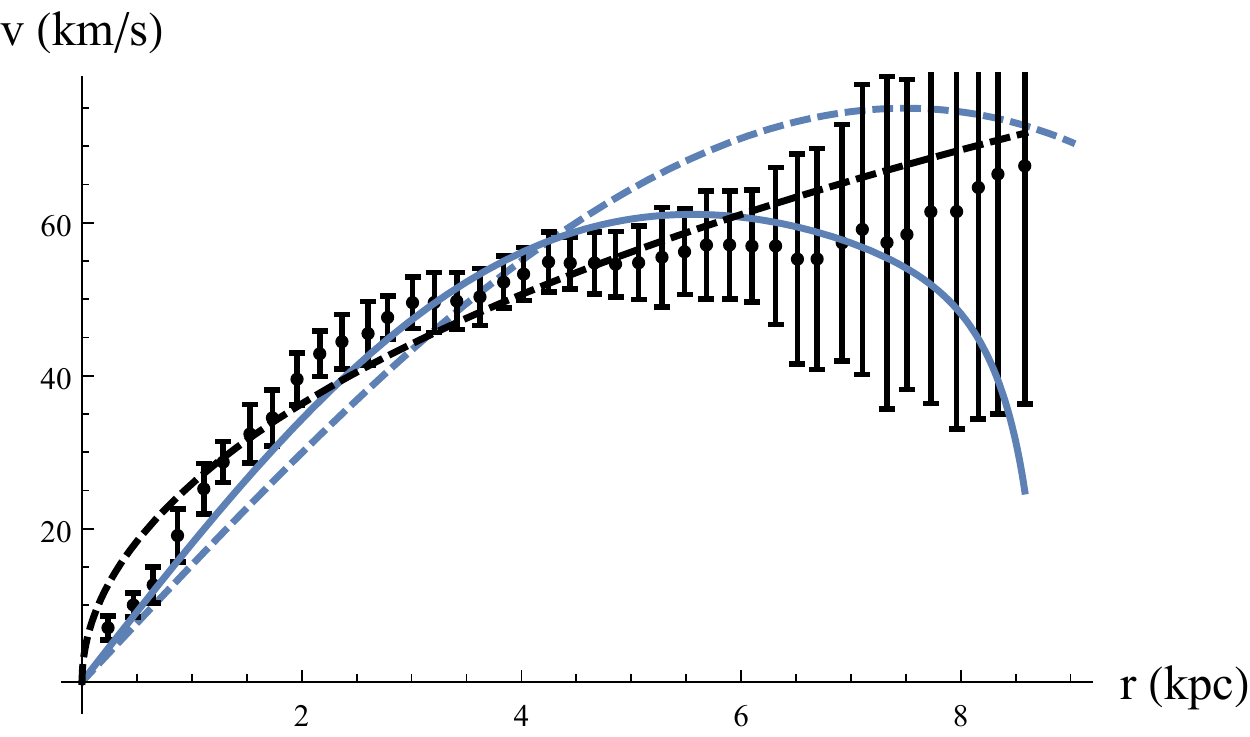}\\ %[2mm]
\vspace{0.5cm} \\ [2mm]
   {\color{white} .}{\hspace{24mm} DDO 39 \hspace{78mm} DDO 154 } \\ %[2mm]
                \vspace{0mm}
 \hspace{8mm}\includegraphics[width=72mm]{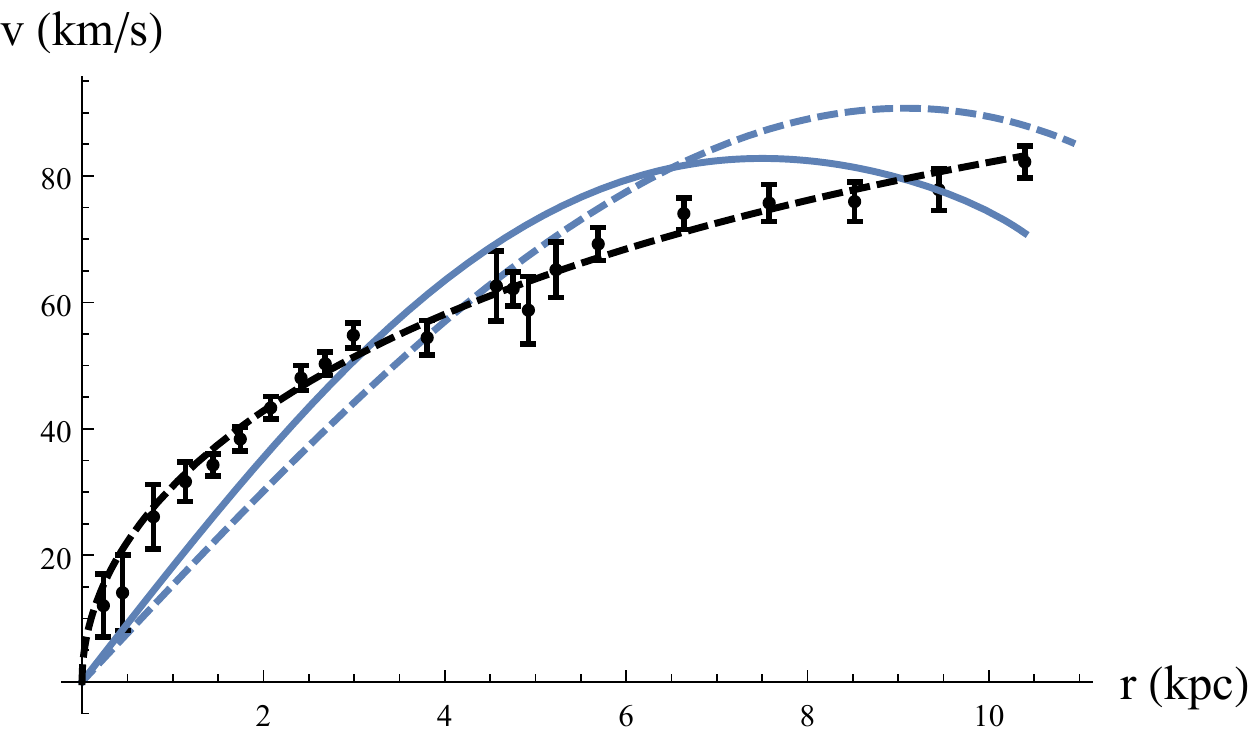} \hspace{25mm}  \includegraphics[width=72mm]{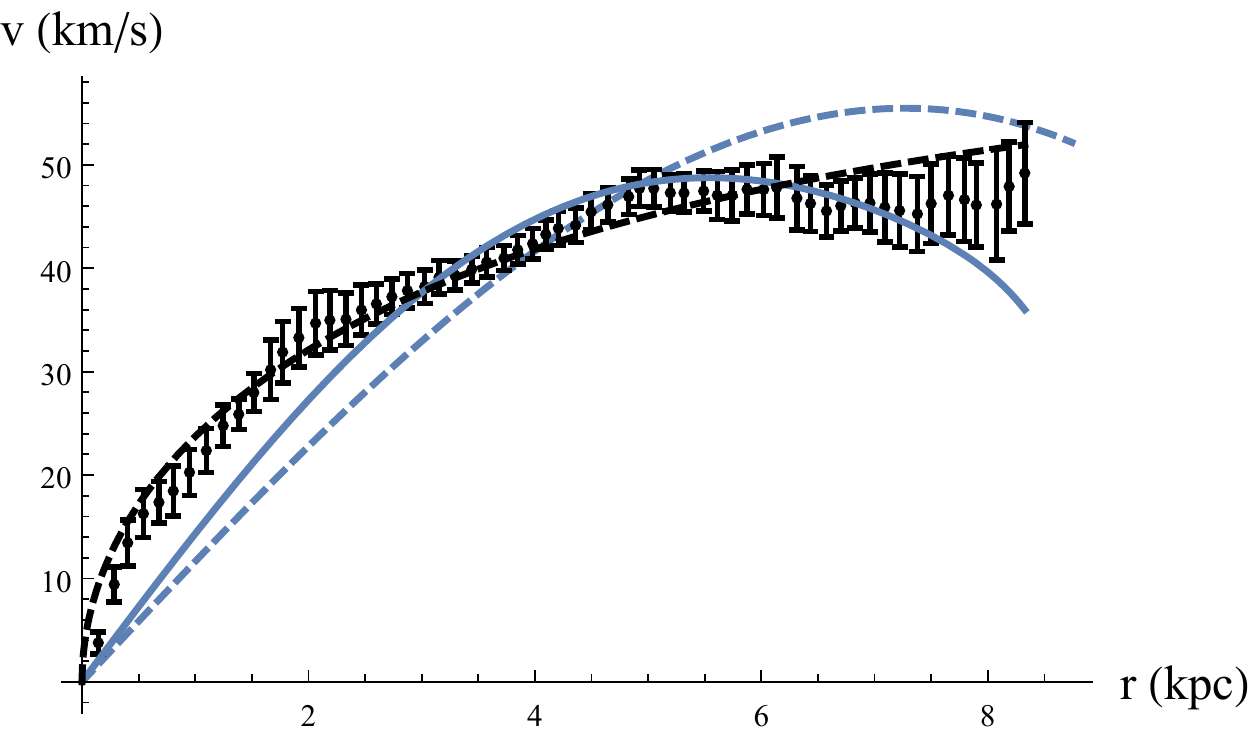} \\ %[2mm]
\vspace{1.5cm} \\
  \end{figure*}%[!]

\end{document}